\documentstyle[preprint]{aastex}

\begin{document}

\title{Cluster AgeS Experiment: The Age and Distance of the Globular
Cluster $\omega$ Centauri Determined from Observations of the Eclipsing
Binary OGLEGC17}

\author{I.~B.~Thompson\altaffilmark{1,2}, J.~Kaluzny\altaffilmark{2,3},
W.~Pych\altaffilmark{3}, G.~Burley\altaffilmark{1}, W.~Krzeminski\altaffilmark{4}, 
B.~Paczynski\altaffilmark{5}, S.~E.~Persson\altaffilmark{1}, 
and G.~W.~Preston\altaffilmark{1}}

\altaffiltext{1}{Carnegie Observatories, 813 Santa Barbara St.,
Pasadena, CA 91101-1292}
\altaffiltext{2}{Visiting Astronomer, Cerro Tololo Inter-American Observatory.
CTIO is operated by AURA, Inc.\ under contract to the National Science
Foundation.}
\altaffiltext{3}{Copernicus Astronomical Center, Bartycka 18,
00-716 Warsaw, Poland}
\altaffiltext{4}{Las Campanas Observatory, Casilla 601, La Serena, Chile}
\altaffiltext{5}{Princeton University Observatory, Department of Astrophysics,
124 Peyton Hall, Princeton, NJ 08554-1001}

\begin{abstract}

We use masses, radii, and luminosities of the detached eclipsing binary
OGLEGC17 derived from photometric and spectroscopic observations to calculate
the age and distance of the globular cluster $\omega$ Cen. Age versus
turnoff mass and age versus luminosity relations from \citet{gira00}
yield two independent estimates of the age, $9.1 < t < 16.7$~Gyr and 
$12.9 < t < 18.5$~Gyr. The distance and distance modulus derived by
use of the infrared versus surface brightness relation are d = $5385\pm300$ pc  and $(m-M)_V=14.06\pm0.11$. Distances derived from our infrared surface
brightness versus color relation and the $Teff$ versus $B-V$ color relation
of \citet{seki00} disagree by about 10 per cent. Major improvements in the
accuracy in estimated age and distance can be made with better 
measurements of the masses of the components of OGLEGC17.

\end{abstract}


\section{Introduction}

Reconciliation of globular cluster ages and the Hubble time is a
primary issue of contemporary observational astronomy. Heretofore, the
single largest uncertainty in the calculation of globular cluster ages
from color-magnitude diagram (CMD) data has been the distance error
\citep{renzini91, vdb96}, followed by the stellar model uncertainties
associated with the choice of the mixing length parameter. The hopes
that Hipparcos \citep{perry97} would resolve the distance issue have
not been fulfilled.  For example, recent determinations of the distance
modulus to 47 Tuc based on Hipparcos subdwarfs gave a value of 13.57
\citep{reid98}, while the value based on Hipparcos red clump giants is
13.32 \citep{jka98}.  The most likely reason for the discrepancy is the
comparison of nearby stars with distant stars, with imperfect
understanding of the population effects.

We avoid these difficulties by measurements of stellar masses and
radii in the eclipsing binary OGLEGC17 in the globular cluster $\omega$
Cen. Detached eclipsing double line spectroscopic binaries offer the
opportunity to calculate age directly from age - turnoff mass
predictions from stellar evolution models.  Distance follows from a
geometrical calculation that employs the Barnes-Evans relation
\citep{barnes76} between color and surface brightness.  The method is
very simple, and is well described by \citet{lacy77}, who has used the
method to measure the distances to a number of well studied systems
\citep{lacy79}. His distance determinations agree well with the recent
trigonometric parallaxes obtained by Hipparcos \citep{sem00}.
\citet{stebbins10} was fully aware of the equivalent method to
determine the surface brightness of the components of Algol, for which
a trigonometric parallax had just been measured. A complete list of
historical references can be found in \citet{krus99}.

There has been steady progress in the last few years in the
empirical calibration of the Barnes-Evans relation with the direct
determinations of angular diameters of many stars  obtained with
optical interferometers \citep{dib98}, and there are excellent
prospects of further improvement \citep{haj98}.  Until recently the
only missing elements have been the detached eclipsing binaries in
globular clusters.  Recent massive photometric searches for the rare
photometric signatures of gravitational microlensing have led to the
discovery of over $ 10^5$ new variable stars.  In particular, the first
detached eclipsing binary near a main sequence turnoff was found in
the globular cluster $\omega$ Cen as a `secondary' project of OGLE
\citep{jka96}.  By now several more such systems have been  found in
several other globular clusters by our group.

Demonstrations of the practical power of using eclipsing binaries for
distance determinations to the Large Magellanic Cloud have been
recently presented by \citet{bell93} (HV 5936), \citet{guinan98} (HV
2274) and \citet{fitz01} (HV 982).

Detached binaries can also free us from problems associated with the
mixing length parameter, because binaries can yield the mass-luminosity
relation, and this is not affected by the mixing length parameter
(\citet{pac97} and references therein).  Also, the helium abundance can
be deduced if the two components are sufficiently different in mass.
This is all possible because detached components are just two single
stars, never affected by the complications of mass exchange, and their
internal structure is not disturbed by the fact that they orbit each
other.  A practical demonstration that all stellar parameters can be
determined with a high accuracy, given accurate photometry and
spectroscopy, is provided by the catalog of well studied detached
eclipsing binaries covering the spectral range from O to M
\citep{and91}.

In this paper we present observations of the detached eclipsing binary
OGLEGC17 in the globular cluster $\omega$ Centauri \citep{jka96}. This
star has a period of 2.467 days. Eclipse depths are  $\Delta$V = 0.30
and $\Delta$V = 0.26 for the  primary and secondary eclipses,
respectively.  With an observed color of $B-V$ = 0.64 and an observed
magnitude of $V$ = 17.16 (out of eclipse), this system lies slightly
above the top of the main sequence in the cluster color-magnitude
diagram, consistent with the individual members of the binary being
stars at or near the cluster turnoff.  These parameters along with the
shape of the light curve suggest that the system is completely
detached, making it ideal to study.  These observations are
used to derive preliminary estimates of the masses of the constituent
stars and the distance and age of the cluster.

In the next section we present a description of the observations and the
reduction analysis. Following that we derive the age and distance of the 
cluster.

\section{Observations}

\subsection{Optical Photometry} \label{opphom}

$B$, $V$ and $I$ light curves were obtained with the $2K \times 2K$
pixel TEK\#5 CCD camera on the du Pont 2.5-m telescope at Las Campanas
Observatory during the period April 21, 1995 through June 4, 1995. The
field of view was $8.9\times 8.9$ arcmin with scale of 0.26
arcsec/pixel.  The individual frames were processed with
IRAF\footnote{IRAF is distributed by National Optical Astronomical
Observatories, operated by the Association of Universities for Research
in Astronomy, Inc., under contract to the National Science Foundation.}
software, and the photometry was measured with DoPhot \citep{sch93}.
Several standard fields from \citet{land92} were observed on 4 nights
during the 1995 run. These observations were used to establish a
transformation between the instrumental $bvi$ and the standard $BVI$
system.  Unfortunately, the only night on which we observed Landolt
standards along with OGLEGC17 turned out to be non-photometric.  The
zero points of our photometry of OGLEGC17 were established based on
data collected during the 1997 observing season.  The same CCD camera
and filters  as for the 1995 observations were used. Instrumental
profile photometry for stars from the cluster sub-field including
OGLEGC17 was derived using the Daophot/Allstar package \citep{stet87}.
Standard stars from \citet{land92} were measured using aperture
photometry while aperture corrections for the cluster field were
derived using the Daogrow program \citep{stet90}.  Based on residuals
observed for standard stars and the uncertainties of the derived
aperture corrections, we estimate that the total errors of the zero
points of our $BVI$ photometry for OGLEGC17 do not exceed 0.02.
Numerous sets of observations taken during the 1997 and 1998 observing
seasons with a CCD camera on the 1-m Swope telescope at Las Campanas
Observatory were used to refine the ephemeris. The final adopted
ephemeris is $T_0(HJD)$ = 2449082.3530 $\pm$ 0.0008 and $P$ = 2.4669384
$\pm$ 0.0000023 days.  The 1995 $BVI$ light curves phased with this
ephemeris are shown in Figure 1. These light curves contain 181, 197
and 101 data points for the $B$, $V$ and $I$-bands, respectively.

On the night of March 16, 1995 (UT) we used the $1K \times 1K$ pixel
TEK\#1 CCD camera on the du Pont telescope to monitor a primary eclipse
of OGLEGC17. The data were reduced in an identical fashion as the time
series photometry described above. The time resolution of these
observations is better than that for the observations collected with
the TEK5 camera.  The derived V-band light curve is shown in Fig. 2.
Note that the eclipse is total,  the phase of constant light
lasts about 0.035P.

Magnitudes and colors observed at maximum light\footnote{As our data
do not cover quadratures we list magnitudes and colors observed at
phase 0.82} and at both minima are listed in Table 1. The totality of
the primary eclipse allows a straightforward determination of the
magnitudes and colors of both components of the binary.  Assuming that
to first order effects due to the ellipsoidality of the components and
to the reflection of light are negligible one may adopt the magnitudes
and colors observed at phase 0.0 (the eclipse of the secondary
component) as corresponding to the larger component of the system.
Note that we have also assumed that there is no "third light" from an
unresolved tertiary component in the OGLEGC17 system. For that star we
obtain $V=17.46$, $B=18.12$, $I=16.63$, $B-V=0.66$ and $V-I=0.83$.  We
adopted the convention that the primary component is the one with
larger size and luminosity.  In this case the secondary component has a
smaller size and luminosity but higher surface brightness than its
companion. That star is eclipsed at  phase 0.0, the deeper minimum.
For the secondary component we obtain $V=18.72$, $B=19.27$, $I=17.915$,
$B-V=0.56$ and $V-I=0.80$.  The position of OGLEGC17 in the cluster
color-magnitude diagram (CMD) is shown in Fig. 3, where the position of
the composite image is shown along with the positions of the individual
stars in this binary. The system consists of a main sequence star
(secondary component) together with a star which has left its
turnoff point (primary component).

\subsection{Infrared Photometry}

Photometric measurements in the $J$, $H$, and $K_s$ passbands were made at
random phases in the period 30 January through 7 February 1999. The
IRCAM infrared camera \citep{sep92} was used on the du Pont telescope
at a scale of 0.348 arcsec/pixel.  Observing, linearization, flat-fielding, and
sky-frame creation methodologies were closely similar to those detailed
in \citet{sep98}.  That paper also presents the definition of the
photometric system, standard stars, and filter transmissions.
Linearized, sky-subtracted, and photometrically calibrated frames were
combined into final stacked images (``mosaics"), which were then
photometered with Daophot. The results are presented in Table
\ref{tbl-2}. The uncertainties returned by Daophot are computed from
the sky level on the raw data frames, and implicitly take the noise to
be Poisson. Measured uncertainties,  computed from the
dispersion in the individual stellar magnitudes, are typically quite
close to the Daophot values, so the latter were adopted and given in
Table 2.

The observations were calibrated with observations of standard stars 
presented in \citet{sep98}. The data for the first three nights were
taken in photometric conditions, and the calibration was determined
independently for these nights. Photometric zero points for the last three
nights were determined by comparing photometry of the other stars in the
field of view to an average of the values for the first three nights.
The $K_s$ photometric system differs negligibly from that of the 
Johnson $K$ system for stars of the relevant spectral type. 
Convolutions of the respective filter transmissions
over Kurucz model atmospheres for such stars show
typical $K - K_s$ values in the range -0.002 to +0.002. 

\subsection{Spectroscopy}

Echelle observations of OGLEGC17 were obtained with the du Pont 2.5-m
telescope between 08 May UT and 13 May UT of 1996. The echelle has a
fixed format optical path feeding a 2D-Frutti detector \citep{shec84}.
The observations were taken with a 1.5 arcsec slit, and the spectral
resolution was about 15 km/sec.  A total of 38 spectra were obtained with
exposure times of 1800 sec. The spectra were extracted with a set of
Fortran programs written at Carnegie Observatories. The final
individual extracted spectra had typical signal levels of between 4 and
6 counts at 4490$\AA$. The extracted spectra were reduced to zero
heliocentric velocity and binned  into nine final spectra according to
the ephemeris in Section \ref{opphom}.

Echelle observations of OGLEGC17 were also obtained with the Blanco 4-m
telescope between 16 April UT and 19 April UT of 1997. Observations
were taken with a 1.5 arcsec slit, and the resulting spectral
resolution was 14 km/sec with the Red CCD camera. The data were
reduced with the echelle package in IRAF.  Exposures were 1800 sec, and
the 22 individual spectra were coadded  into four final spectra
according to the ephemeris in Section 2.1.

Multiple observations of the star HD193901 ($B-V= 0.53$, [Fe/H] = -1.2
\citep{tomkin}) were obtained at each telescope and reduced with the
appropriate software.  The individual spectra were averaged to produce
a high signal-to-noise template for the velocity measurements.  The
template spectrum was rotationally broadened to $V_{rot}$ = 20 km/sec.
Adopting the radii from Section 3, and assuming that the
components are rotationally locked, leads to an expected rotational
velocities of 38 km/sec for the primary and 18 km/sec for the
secondary.

Velocities were measured with the IRAF routine FXCOR. The results are
presented in Table \ref{tbl-3} where we give the number of spectra
averaged at each phase point, the mean phase, the total range in phase,
and the velocities and the errors returned from FXCOR. As demonstrated
by \citet{lath88}, all halo binaries with periods less than about ten
days have circular orbits. From this, and because of  the relative
quality of our radial velocity measurements, we assume that OGLEGC17
has a circular orbit (see also the discussion by \citet{lucy71}). We
made a least squares fit to the velocity data using GaussFit, solving
for the systemic velocity $\gamma$ and the velocity amplitudes $K_1$
and $K_2$. The results of this fit are presented in Table 4.  The
measured velocities together with the fit are presented in Figure 4.
The measured value of $\gamma$ (237.97~$\pm$~1.93 km/sec) confirms that
OGLEGC17 is a member of the cluster, the systemic velocity of $\omega$
Cen is 232.3 km/sec \citep{harris96}. Adopting an inclination of $i$ =
86.3 from the photometric solution in Section 3, we derive $M_1$
=0.806$\pm$0.056~$M_{\odot}$ and $M_2$ =0.686$\pm$0.047~$M_{\odot}$.

\section{Photometric solution}

The light curves were analysed by use of the Wilson-Devinney (1971; hereafter
WD) model as implemented in the 1986 version of the code. The code  is
described in some detail by Wilson (1979) and by Leung \& Wilson
(1977).  The MINGA minimization package (Plewa 1988).\footnote{ The
MINGA package can be obtained from http://www.camk.edu.pl/$\sim$plewa} was
used for the actual fitting of the observed light curves and the
derivation of system parameters.  The bolometric albedo and gravity
brightening coefficients were set to values appropriate for stars with
convective envelopes: $A1=A2=0.5$, $g1=g2=0.32$. The mass ratio was
fixed to the spectroscopic value $q=m2/m1=0.851$. We note at this point
that because of the well detached configuration of the binary the
modeled light curves show only a very small dependence on the assumed
value of $q$.  Hence, the uncertainty in $q$, amounting to 0.073,
has little effect on the photometric solutions presented below. In addition,
uncertainties in the adopted values of the effective temperatures of
the components do not  noticeably influence the results of our analysis.
These temperatures were estimated using the empirical calibration
$T_{eff}=T_{eff}(g,(B-V)_{0},{\rm [Fe/H]})$ published by Sekiguchi \&
Fukugita (2000).  The observed colors were corrected for interstellar
reddening assuming $E(B-V)=0.13$ as derived for the coordinates of
OGLEGC17 from maps of Schlegel et al. (1998).  We adopted ${\rm
[Fe/H]}=-1.7$ for the metallicity of OGLEGC17 (see Sec. 4 for
discussion of the reddening and the metallicity of OGLEGC17).
Theoretical linear limb-darkening coefficients in the $BVI$ bands were
taken from Claret, Diaz-Cordoves \& Gimenez (1995) and from
Diaz-Cordoves, Claret \& Gimenez (1995) for the adopted values of
effective temperature and metallicity.

We assumed a  detached configuration for OGLEGC17 and analysed the
light curve by running the WD code in both modes 0 and 2.  as described
in detail by Leung \& Wilson (1977).  In Mode 0 the luminosities and
temperatures of the components are not coupled. The adjustable
parameters are the inclination $i$,  the dimension-less potentials
$\Omega_{1}$ and $\Omega_{2}$, and the relative luminosities $L_{1}$
and $L_{2}$.  The three light curves were analysed simultaneously so the
total number of adjustable parameters was equal to nine.  In Mode 2 the
luminosities are coupled to the temperatures. For that mode the
adjustable parameters are $i$, $\Omega_{1}$ and $\Omega_{2}$, the
temperature of the secondary component $T_{2}$, and the luminosity
$L_{1}$ (a total of 7 free parameters).  For both modes  stable
solutions were derived and the resulting parameters and formal errors
are listed in Table 5.

The parameters obtained from both solutions are consistent with each
other to within the formal uncertainties. The solution corresponding to Mode 2
shows slightly smaller errors for most parameters and given that this
solution has fewer free parameters we adopt it in the following
discussion. In Fig. 5 we show the residuals of the fit along  with the
synthetic light curve overlaid on the observed light curves. In Table 6
we list some of the absolute parameters of the system obtained by combining
data from Tables 4 and 5.

The results of our analysis show that OGLEGC17 is a well-detached
system. Its primary exhibits a slight ellipsoidality with
$r_{pole}/r_{point}=0.976$.\footnote{ $r_{pole}$ is the "polar" radius
while $r_{point}$ is measured in the direction toward the mass center
of the binary.} The secondary component is almost spherical with
$r_{pole}/r_{point}=0.997$.  As noted above, the primary is slightly
evolved, with a larger total luminosity and lower surface brightness
than the secondary.

\section{Age determination}

The fact that the primary component of OGLEGC17 has already left the
cluster turnoff should allow a precise determination of its age by
use of  age-luminosity and age-turnoff mass relations derived from
models of stellar structure and evolution. We adopt the models of
\citet{gira00} in the following discussion.

The two  main sources of uncertainty in our determination of the age of
OGLEGC17 are the relatively large errors in the estimated masses of
both components and  the metallicity of this particular
system.  It has been known for many years that unlike other globular
clusters in our Galaxy $\omega$~Cen contains stars with a range of
metallicities (Norris \& Bessell 1975). Suntzeff \& Kraft (1996) used a
sample of 379 stars (234 and  145 from the lower and upper giant
branch, respectively) with spectroscopically derived [Fe/H] to study
the metallicity distribution in the cluster.  They found that all but a
few stars have [Fe/H] ranging from $-1.9$ to $-0.6$\footnote{ We are
using the Zinn-West scale of [Fe/H] throughout.}.  The observed
distribution rises rapidly at ${\rm [Fe/H]}=-1.85$ with a median value
${\rm [Fe/H]}=-1.7$. In addition, 78\% of their sample has $-1.85<{\rm
[Fe/H]}<-1.4$. A recent study by Majewski et al. (2000) shows quite
convincingly that the metallicity distribution of $\omega$~Cen giants
terminates at the metal-rich end at $[{\rm Fe/H}]\approx -1.15$.  The
apparent discrepancy with  earlier studies is explained by more a
careful elimination of field interlopers in the Majewski et al. study.

Unfortunately, our spectra have very low $S/N$ ratio and cannot be used
for a direct determination of the system metallicity. However, we can
use the location of the components of OGLEGC17 on the cluster CMD (see
Fig. 3) to estimate  metallicity.  Several authors have suggested that
$\omega$~Cen harbors  a few sub-populations of stars, not only of
different metallicities but also of different ages (Lee at al. 1999;
Hughes \& Wallerstein 2000; see also references therein). It has been
estimated that ages of the cluster stars cover a range of about 3~Gyr
with the more metal rich stars being younger.  If this is true then the
position of the primary component in the cluster CMD is not very useful
as effects of age and metallicity cannot be easily deconvolved at the
top of the main-sequence - for a fixed age stars with higher metallicity
are fainter, but for fixed [Fe/H] younger stars are brighter. Since the
primary component is on the bright side of the "clump" of stars leaving
the main-sequence it can be either old and metal poor, or young and metal
rich.  The secondary component of the binary is located on the blue
edge of the upper main-sequence, just below the turn-off.  Hence 
its metallicity places it among the bulk of $\omega$~Cen
stars with ${\rm [Fe/H]}\approx -1.7$.

In the following  discussion we will derive results for two assumed values
of metallicity: ${\rm [Fe/H]}=-1.74$ and ${\rm [Fe/H]}=-1.33$.  The
lower value is marginally smaller than the median value for the bulk of
the cluster giants (Suntzeff \& Kraft 1996) and the higher one marks
the lower end of the observed metallicity distribution. In addition,
modern evolutionary tracks suitable for our analysis are available in
the literature for these two values of [Fe/H] (see below).

\subsection{Age - Turnoff Mass Relations}

The most straightforward way to estimate the age of the cluster is to
use the fact that the mass of the primary star in OGLEGC17 differs
negligibly from the turnoff mass of $\omega$ Cen.  The time it takes a
main sequence star to exhaust its core hydrogen is a monotonic function
of mass therefore our estimate of the mass of the primary star derived
from spectroscopic and photometric data can be used to directly measure
the age of the cluster.

Figure 6 shows age - turnoff mass relations from \citet{gira00} for metallicities Z =
0.0004 and Z = 0.001, corresponding to ${\rm [Fe/H]}=-1.74$ and ${\rm
[Fe/H]}=-1.33$, respectively, assuming $logZ=0.977{\rm [Fe/H]-1.699}$
\citep{bert94}. These values bracket the likely value for the
metallicity of OGLEGC17. The solid line and dotted lines represent the
mass of the primary star and one sigma errors, $0.806\pm0.056
M_{\odot}$. The resulting ranges in the age of the primary of OGLEGC17
are $9.1<t_{1}<15.8$~Gyr for Z = 0.0004, and $9.5<t_{1}<16.7$~Gyr for Z
= 0.001. A formal error in the estimated age of $\pm$1 Gyr could be
obtained by reducing the uncertainties in the masses to $\pm 0.015 M_{\odot}$.
The corresponding  uncertainty in the $K_{1,2}$ velocities is $\pm$1 km/s.

\subsection{Age - Luminosity Relations}

We can also estimate the age of the cluster by using age versus
luminosity relations. An advantage of this approach is that the
parameters of the secondary component provide an independent constraint
on the age of the binary and therefore can be used for testing and/or
improving the result derived from analysis of the primary component
alone.  To minimize the uncertainties involved in a comparison of
observed quantities with calibrations provided by models we use the age
versus bolometric luminosity relation \citep{gira00}. Bolometric
luminosities from models are unaffected by uncertainties associated
with model isochrones relating $T_{eff}$ and $L_{bol}$ to color index
and  absolute magnitude in a selected band.  First we have to determine
bolometric luminosities for both components.  We calculate these from
the standard relation $L_{bol}= 4 \pi R^{2} \sigma T_{eff}^4$.  Two
recent empirical calibrations of $T_{eff}=T_{eff}(g,[Fe/H],B-V)$ have
been published by Alonso et al.  (1996) and by Sekiguchi \& Fukugita
(2000; SF hereafter).  These are in good agreement with each other. For
the range of colors and metallicities discussed below,  the calibration
by Alonso et al. (1996) gives values of $T_{eff}$ which are
systematically lower by  25-30 deg than the $T_{eff}$ based on the SF
calibration.  That difference  has only a small impact on calculated
$L_{bol}$ (as compared with errors associated with uncertainties  of
radii, $E(B-V)$ and $B-V$) and therefore we will use values of
$T_{eff}$ based on the SF calibration.  To estimate $T_{eff}$ we need
to know both  $(B-V)_{0}$ and [Fe/H].

The Harris catalogue (Harris 1996) gives $E(B-V)=0.12$ for
$\omega$~Cen.  According to maps of the extinction by Schlegel et al
(1998) $E(B-V)$ varies by about 0.02 across the field of the cluster,
with the reddening slightly larger in the southern part of the
cluster.  For the position of OGLEGC17 we get $E(B-V)=0.132$. We will
adopt for our analysis $E(B-V)=0.13\pm 0.02$.  Assuming 0.02 as an
error of our $B-V$ measurements we have $(B-V)_{0,1}=0.53 \pm 0.03$ and
$(B-V)_{0,2}=0.43 \pm 0.03$.

We also calculate the bolometric luminosities of the components of
OGLEGC17 using calibrations of $T_{eff}=T_{eff}([Fe/H],V-K)$ from
Alonso et al. (1996, 1999) for dwarf and giant stars. As discussed in
Section 5.2, we have to assume values  for the luminosity ratios
for the two components in OGLEGC17 because we do not have infrared
observations of the eclipse profiles. Adopting the luminosity ratios
given in Section 5.2, and assuming a reddening of $E(B-V)=0.13\pm 0.02$,
we derive $(V-K)_0$ = 1.40 for the primary and $(V-K)_0$ = 1.17 for the
secondary.

In Table 7 we list the effective temperatures derived for both
components of OGLEGC17 using the SF and Alonso et al. calibrations for
metallicities Z = 0.0004 and Z = 0.001. For each component we list the
formal errors in $T_{eff}$ including  a 0.03 uncertainty
$(B-V)_{0}$ as well as an  rms error from the color - effective
temperature calibration.  These values of $T_{eff}$ and the absolute
radii listed in Table 6 are used to calculate the bolometric
luminosities. The resulting values of $L_{bol}$ along with their formal
errors are  given in the last two columns of Table 7.

The luminosity of a star of a given mass is a function of 3
parameters:  age $t$, initial metallicity $Z$ and initial helium
abundance $Y$.  While $Z$ (or more precisely [Fe/H]) can be relatively
easily determined from the observations, the helium abundance cannot
measured directly in main sequence stars of globular clusters.
In the case of a binary star whose components
have the same initial composition there is, however, some redundancy of
information. Hence, knowing the masses of both components and [Fe/H] we
may determine both $Y$ and $t$ from the observed luminosities
\footnote{To be precise, such a procedure is applicable only to
detached systems which have not undergone any  mass transfer episodes
during their evolution}. A more thorough discussion of this subject is
given by Paczynski (1997).

In Figure 7 we show age versus luminosity relations based on
evolutionary tracks recently published by Girardi et al. (2000). The
left panel is for models with $[Z=0.0004,Y=0.23]$  while the right one
is for $[Z=0.001,Y=0.23]$.  In each panel we show relations for the
primary star mass $\pm$ one sigma errors ($m_{1}=0.806\pm
0.056~M_{\odot}$ with dotted lines and  relations for the secondary
mass $\pm$ one sigma errors ($m_{2}=0.686\pm 0.047~M_{\odot}$) with
solid lines. These relations were derived from evolutionary tracks for
$m/m_{\odot}={0.6,0.7,0.8,0.9}$, interpolating to the appropriate mass
using a method developed by Weiss and Schlattl (2000). Vertical lines
in Figure 7 mark $L\pm\sigma_{L}$ ranges for $L_{1}$ and $L_{2}$.  The
intersections of age - luminosity relations with lines marking 1 sigma
limits on $L_{1}$ and $L_{2}$ give limits on the age for a given mass.
For the $B-V$ effective temperature calibration and  $Z=0.0004$ we
obtain $9.1<t_{1}<17.0$~Gyr  for the primary component of OGLEGC17. For
the secondary component we adopt a conservative lower limit to the age
corresponding to the age at  the lower one sigma values of mass and
luminosity:  $t_{2}>12.9$~Gyr. This leads to a plausible age of the
binary $12.9<t<17.0$~Gyr.  Similarly for $Z=0.001$ we obtain
$9.8<t_{1}<18.4$~Gyr and $t_{2}>14.9$~Gyr leading to an age of
$14.9<t<18.4$~Gyr.  For the $V-K$ effective temperature calibration
and  $Z=0.0004$ we obtain $9.6<t_{1}<17.1$~Gyr for the primary
component. For the secondary component we find $t_{2}>15.0$~Gyr. This
leads to a plausible age of the binary $15.0<t<17.1$~Gyr.  Similarly
for $Z=0.001$ we obtain $10.4<t_{1}<18.5$~Gyr and $t_{2}>16.4$~Gyr
leading to an age of $16.4<t<18.5$~Gyr.

Our limits on the age of the primary do not depend strongly on
metallicity or which color is used to determine the effective
temperatures. In all cases, the derived age of the secondary reduces
the limits on the ages. We conclude that for the masses and
luminosities derived for both components  of OGLEGC17 the age  of the
binary can be constrained to the approximate range $12.9<t<18.5$~Gyr.

We note that accurate determination of $t_{2}$ would require narrowing
of errors for both $m_{2}$ and $L_{2}$.  However, an improved estimate
of $t_{1}$ is possible just by narrowing a range of allowable values of
$m_{1}$.  This is is due to fact that the primary component has already
reached an evolutionary stage at which luminosity is a steep function of
age. Within the formalism of our age estimate, a reduction in the error
in the age to $\pm1.0$~Gyr would require improving the error in the $K$
velocities to approximately $\pm0.5$ km/sec, a value that should be
easily achievable with observations of OGLEGC17 with the new generation
of echelle spectrographs on southern large telescopes (eg. Torres et
al. 1997; Metcalfe et al. 1996). These data would lead to accuracies in
the masses and absolute radii of better than one per cent. In addition,
the systematic error associated with the uncertain metallicity of this
system can be addressed by high signal-to-noise echelle observations
at quadrature, where the relative velocities of the two components are
large enough to allow an  metallicity estimate for each of the
components.

\subsection{Dependence of derived ages on stellar models}

The limits on the age of OGLEGC17 listed above were derived using
stellar tracks taken from models published by \citet{gira00}.
One may wonder to what extend our results depend on the selection of a
particular set of stellar models.  A detailed comparison of different
sets of models is beyond the scope of this paper. However, we performed a
limited test by comparing age-luminosity relations from Girardi  et al.
(2000) with relations constructed from tabular data given by Weiss \&
Schlattl (2000). Analytical expressions given by Weiss \& Schlattl
allow one to construct age-luminosity relations for a large range of
masses and chemical composition defined by ${Z,Y}$. Hence, given masses
and luminosities for the components of a given binary, one may derive
not only the age but also the helium content $Y$.  Age-luminosity
relations were extracted from two sets of models for
$[Z,Y]=[0.0004,0.23]$ and for $[Z,Y]=[0.001,023]$ and for masses
$0.7_{M\odot}$ to $0.8_{M\odot}$.  A comparison of these two sets of
models is presented in Figure 8 which shows differential residuals
$\delta t$ versus luminosity for models with $[Z,Y]=[0.0004,0.23]$ and
for two values of the mass.  We conclude that relations based on these
two recent sets of models are in very good agreement with each other
for the considered range of parameters.

\section{Distance to the cluster}

OGLEGC17 offers the opportunity for determining the distance to a
globular cluster in a way independent of other measurement
techniques. Here we present two estimates, one based on the bolometric
luminosities of the components of the binary, and the second based on
relations between  the surface brightness and  infrared colors.

\subsection{Bolometric Luminosities}

We may use the bolometric luminosities listed in Table 7 to calculate
the absolute visual magnitudes $M_{\rm V}$ of both components of the
binary. Bolometric corrections were obtained by interpolating
tabular data given in \citep{houd00}  for unreddened color
indices $(B-V)_{0,1}=0.53$ and $(B-V)_{0,2}=0.43$ and gravities
$log~g_{1}=3.8$ and $log~g_{2}=4.4$. We obtained bolometric corrections
$BC_{1}=-0.22 $ and $BC_{2}= -0.21 $ for ${\rm [Fe/H]}=-1.74]$ and
$BC_{1}=-0.20 $ and $BC_{2}=-0.19 $ for ${\rm [Fe/H]}=-1.33$. Our
formal uncertainties of $(B-V)_0$ of 0.03 transform into formal
errors for bolometric corrections of about 0.01. Adopting
$M_{bol\odot}=4.70$ (eg. Salaris \& Weiss 1988) and using the bolometric
luminosities from Table 7 we arrive at the absolute visual magnitudes
given in Table 8. We then use the apparent $V$ magnitudes of both
components to calculate the distance moduli and these are also given in
Table 8. Values of $(m-M)_{V}$ derived for the primary and the
secondary component are consistent with each other.  For the $(B-V)$
effective temperature calibration and ${\rm [Fe/H]}=-1.74$ we obtain an
average value $(m-M)_{\rm V}=13.78\pm0.16$ while for ${\rm
[Fe/H]}=-1.33$ we have $(m-M)_{\rm V}=13.86\pm0.16$. For the $(V-K)$
effective temperature calibration and ${\rm [Fe/H]}=-1.74$ we obtain
an average value of $(m-M)_{\rm V}=14.00\pm0.09$ and for 
${\rm [Fe/H]}=-1.33$ we obtain $(m-M)_{\rm V}=14.00\pm0.09$.

These values are broadly consistent with $(m-M)_{\rm V}=13.84$ adopted
by Harris (1996)  for $\omega$ Cen.

\subsection{Surface Brightnesses}

We estimate the distance to OGLEGC17 by use of empirically
calibrated relations between surface brightness and infrared color.  Di
Benedetto (1998) presented a calibration of  $S_{\rm V}$ vs $(V-K)$ for
a sample of nearby, Population I stars with luminosity classes III, IV,
and V.  We have extended his calibration to $V-I, V-J$, and $V-H$ for
the same set of stars.

Figure 9 presents these photometric data, together with quadratic
least-squares fits. Values for $S_V$ and for $(V-K)_0$ were taken from
Di Benedetto (1998), and the rest of the photometric data  were taken
from archival tables accessed at the Centre de Donn\'ees astronomiques de
Strasbourg. Note that the calibrating photometry
is in the Johnson system. The terms for the quadratic fits are
presented in Table 9 together with the rms values of the fits (some
obviously discrepant points in Figure 9 were not included in these
fits). Figure 9 also shows reddening vectors for a reddening of $A_V =
1.0$. The vectors are essentially parallel to the relations,
indicating that modest errors in the reddening do not impact the
estimates of the physical distance to the binary (note, however, that
errors in the reddening do directly affect the final estimates of the
distance modulus $(m-M)_V$). Given the relative rms values of the fits,
and the steepness of the $S_V-(V-I)$ relation, we estimate the distance
to OGLEGC17 using only our $J, H$, and $K$-band data.

The stars used in the surface brightness calibration are of necessity
in the solar neighbourhood, and therefore most likely  are Population I
stars. To see how the surface brightness versus color calibration depends on
metallicity we have used synthetic color-temperature relations from
\citet{houd00}.  Figure 10 presents $S_V$ - color relations extracted
from Table 5 in \citet{houd00} for [Fe/H] = 0.0 and [Fe/H] = -2.0. As
also found by Di Benedetto (1998), these relations show little
sensitivity to metallicity, the correction amounting to approximately
+0.04 in $S_V$ for the $J$ filter and +0.02 in $S_V$ for the
$H$ and $K$ filters.

Table 2 presents mean magnitudes for $J, H,$ and $K$-band observations
of OGLEGC17 taken out of eclipse. Assuming $V = 17.155$ for OGLEGC17 at
quadrature, and assuming a reddening of $E(B-V) = $ 0.13 and a
reddening law from \citet{rieke85}  we calculate the mean colors for
the system as a whole listed in column 2 of Table 10. The WD solution
for the light curve (see Figure 5) indicates that the ellipticity of
the primary and, to a lesser extent, reflection effects in the binary,
produce a one percent modulation of the $V$-band light out of
eclipse. We have ignored these effects in calculating mean colors. In
order to calculate the colors of the individual components we need to
know the relative luminosities of the primary and secondary. These
ratios require light curves through the eclipses in the infrared bands,
data which we do not have. The WD code used to estimate these ratios by
extrapolating the $B, V$, and $I$-band solutions to $J, H$, and $K$
(Mode 2, see Table 5). These values are $q_J = 3.922, q_H = 4.038$, and
$q_K = 4.136$ with estimated errors of $\pm0.05$. The unreddened colors
of the individual components of the binary are listed in columns 3 and
4 of Table 10.

The surface brightness $S$ of a star is defined as: 
\begin{eqnarray}
m_{0}=S-5log \phi 
\end{eqnarray} 
where $m_{0}$ is the un-reddened apparent magnitude of a star of
angular diameter $\phi$. The zero-point is set such that $S=m_{0}$ for
$\phi=1$ mas. From Di Benedetto, equation 4, we have 
\begin{eqnarray}
\phi(mas) = 10^{(S_{V} - 0.2*(V - A_{V}))},
\end{eqnarray}
and from \citet{lacy77}, equation 4, we have 
\begin{eqnarray}
d(pc) = 1.337 \times 10^{-5} r(km)/\phi(mas).
\end{eqnarray}
Also
\begin{eqnarray}
(m-M)_V = 5.0 * log(d(pc)) - 5.0 + A_V.
\end{eqnarray}

We use colors from Table 10 (transformed to the Johnson system using
relations in \citet{sep98} and \citet{bessel88}) together with the
relations in Table 9 to calculate the surface brightnesses of the
components of the binary; these are listed in columns 2 and 3 in Table
11. The above equations are then used to calculate the distances
(columns 4 and 5 in Table 11) and the distance moduli (columns 6 and 7
in Table 11).

We adopt a distance which is the average of the values derived for the
$S_V$ estimates from $V-J$, $V-H$, and $V-K$ for the primary and
secondary, all values weighted equally. The final values are $d=5469$
pc and $(m-M)_V=14.09$. To estimate the errors in these values we
determined the sensitivity of the answers to one sigma variations in
the input observations. The physical distance changes 4.0 per cent for
a one sigma change in $K_{1}$ and $K_2$, 2.9 per cent  for
$r_1/(a_1+a_2)$ and $r_2/(a_1+a_2)$, 2.2 per cent for $J$, $H$, and $K$
photometry, 0.8 per cent for $V$ photometry, and 0.7 per cent  for
$E(B-V)$. The result is insensitive (0.3 per cent change) to variations
in $q_V$, $q_J$, $q_H$, and $q_K$.  The quadrature sum of these
percentage errors totals 5.5 per cent, and our final estimate of the
distance to $\omega$ Centauri is $5385\pm300$ pc. The distance modulus is
$(m-M)_V=14.06\pm0.11$. Note that the error in the reddening
is much more important in determining the distance modulus. It is interesting
to note that the values for the distance modulus derived from our
$K$-band observations (through bolometric
luminosities using effective temperatures calibrated with $V-K$ colors and
through surface brightnesses estimated from $V-J$, $V-H$, and $V-K$ colors)
are very similar and differ at the level of 0.20 magnitudes from the
other estimates of the distance modulus of $\omega$ Cen.

\subsection{Discussion}

The error in the estimation of the physical distance is dominated by
the accuracy of the velocity curve and the accuracy of the photometric
solution for the radii. As mentioned in the previous section, it should
be possible to lower the errors in the radial velocity curve by at
least a factor of five by use of improved  echelle observations of OGLEGC17.
In addition, detailed visual and infrared observations of the light
curves will improve the photometric solution for relative radii and
luminosities. Given that the primary eclipse is total and that the
binary is detached, a precision of the order of 0.1\% is attainable for
these relative parameters. In addition, we will be able to replace the
estimates of the effective temperatures based on $(B-V)$ with superior
estimates based on infrared colors, significantly reducing errors
arising from uncertain metallicity of OGLEGC17.

\section{Summary}

We have presented photometric and spectroscopic observations of the
detached, eclipsing binary OGLEGC17 in the globular cluster $\omega$
Cen.  We have used these data to estimate the age of this system to be
$9.1<t<16.7$~Gyr from age - turnoff mass relations and
$12.9<t<18.5$~Gyr from age - luminosity relations. We have used an
empirical calibration of surface brightness with infrared color to
estimate a distance of $5385\pm300$ pc to the cluster, a distance
modulus of $(m-M)_V=14.06 \pm 0.11$. This is in disagreement with an
estimate of $(m-M)_V=13.79$ based on the measured bolometric
luminosities of the components of the binary.

These estimates can be improved in a straightforward way by further
observations, in particular better determinations of the $K_1, K_2$
velocities and of the visual and infrared eclipse light profiles.
Formal errors in the age of $\pm1$~Gyr are achievable by reducing the
errors in the velocity curves to $\pm$1~km/s.

\acknowledgments

JK and WP were supported by the Polish Committee of Scientific Research
through grant 2P03D003.17 and by NSF grant AST-9819787 to Bohdan
Paczy\'nski. IT, SEP and GWP  were supported by NSF grant AST-9819786.
JK thanks Robert Lupton for his continuing interest in the progress of
this work.

\clearpage
\begin{figure}
\figurenum{1}
\plotone{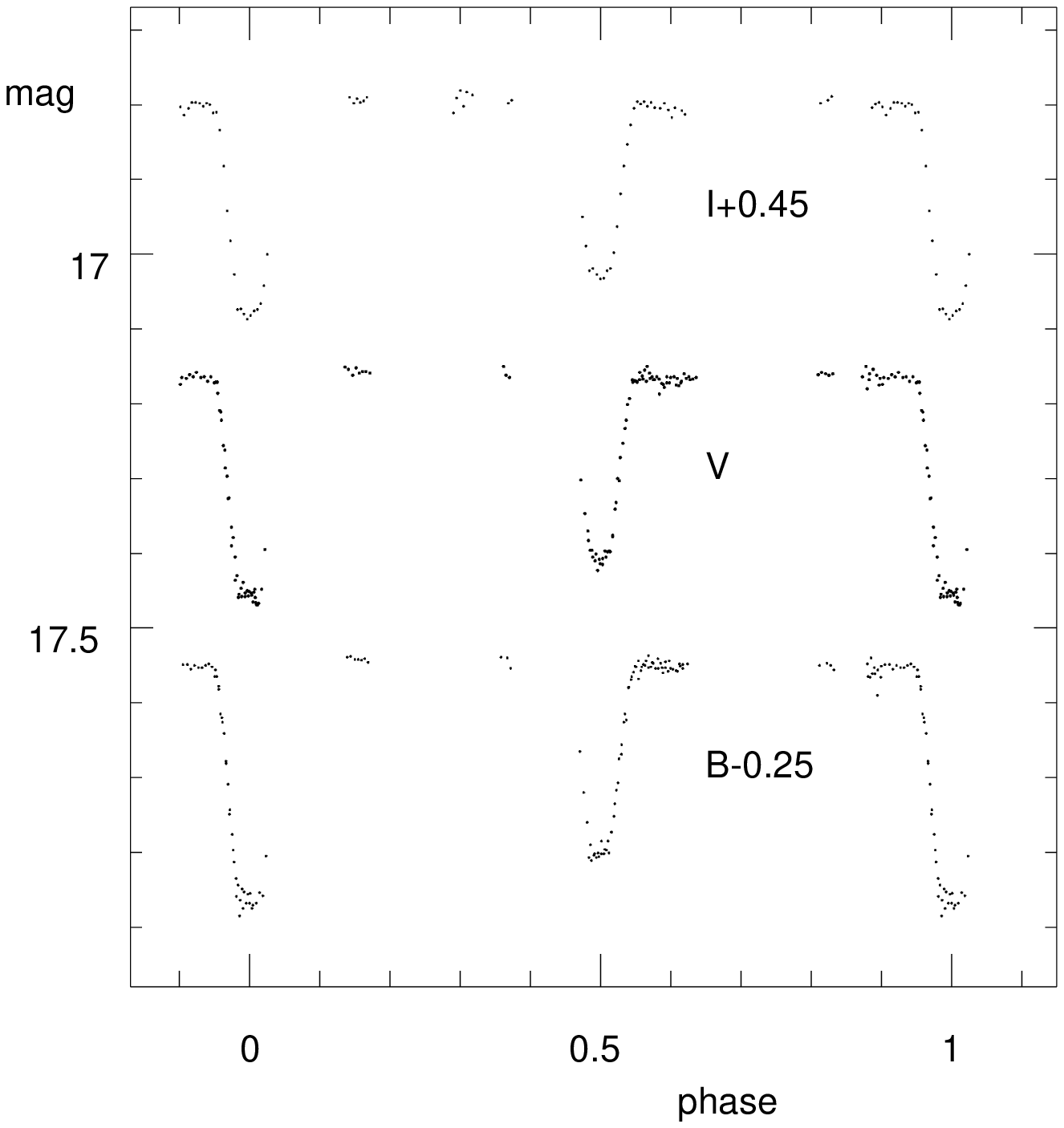}
\caption{$BVI$ light curves of OGLEGC17. \label{fig1}}
\end{figure}
\clearpage
\begin{figure}
\figurenum{2}
\plotone{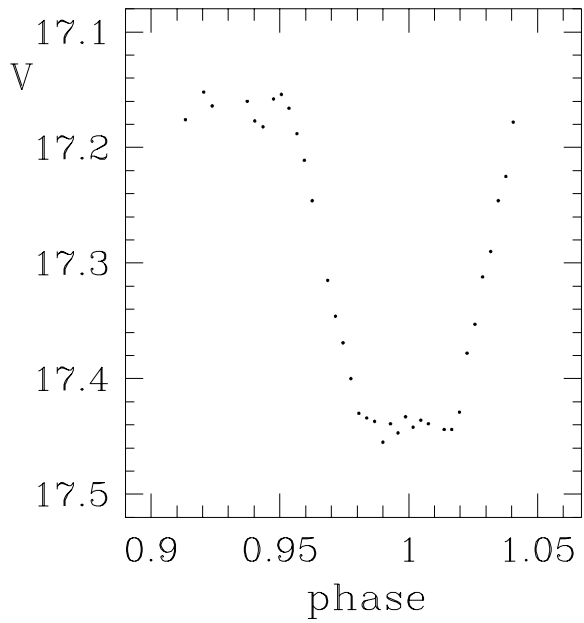}
\caption{The $V$-band light curve of OGLEGC17 obtained with the TEK1
camera.  \label{fig2}}
\end{figure}
\clearpage
\begin{figure}
\figurenum{3}
\plotone{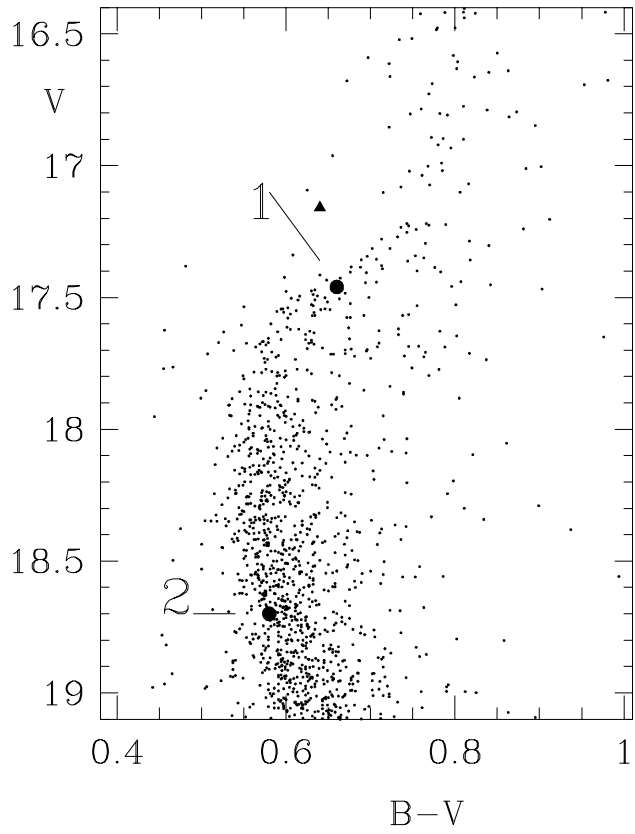}
\caption{The $\omega$ Cen CMD for stars near OGLEGC17. The locations of
both components of the binary are marked along with the position of the
composite image (filled triangle).  \label{fig3}}
\end{figure}
\clearpage
\begin{figure}
\figurenum{4}
\plotone{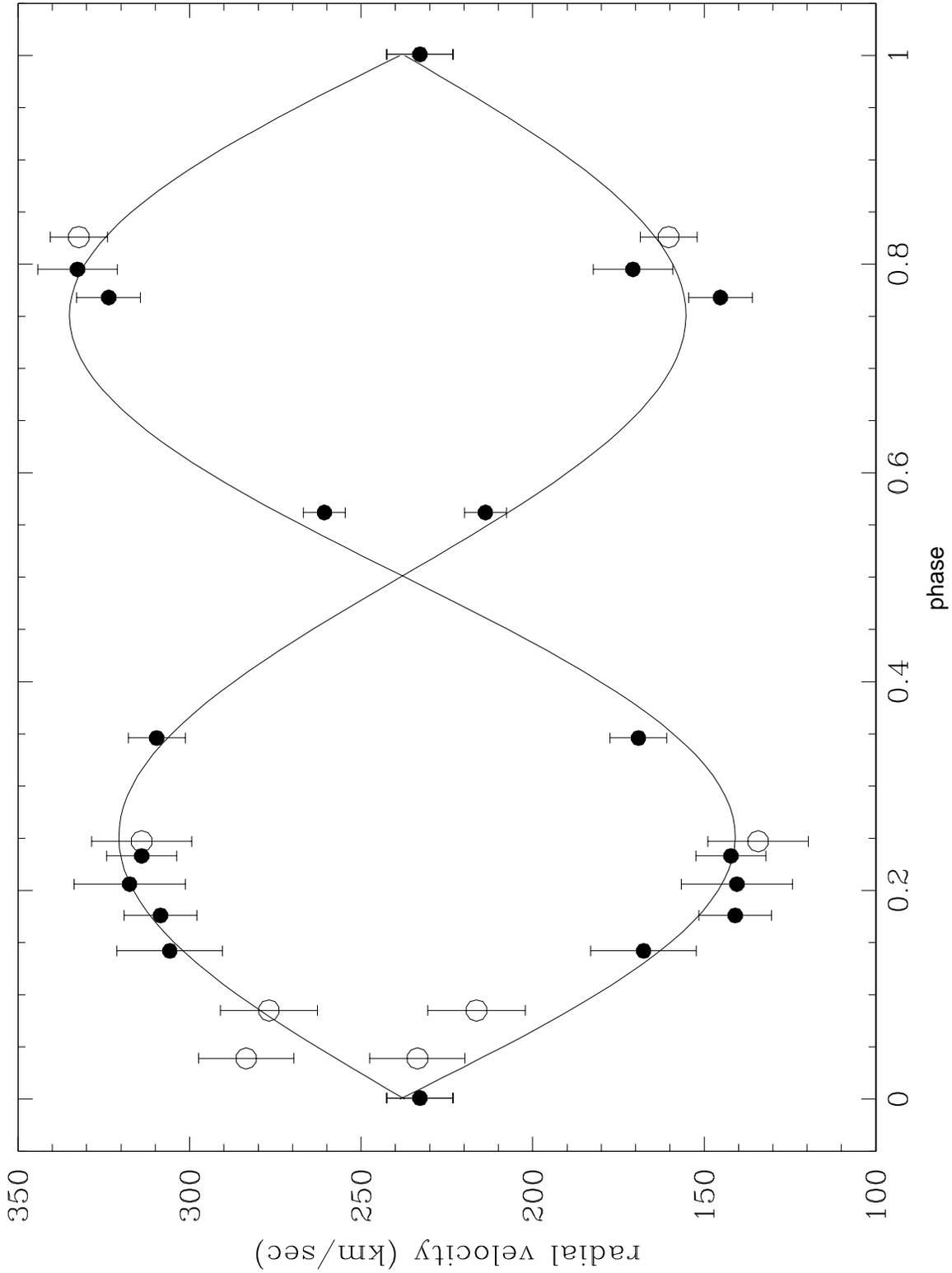}
\caption{The LCO (filled symbols) and CTIO (open symbols) radial
velocities plotted with the GaussFit orbital solution.  \label{fig4}}
\end{figure}
\clearpage
\begin{figure}
\figurenum{5}
\plotone{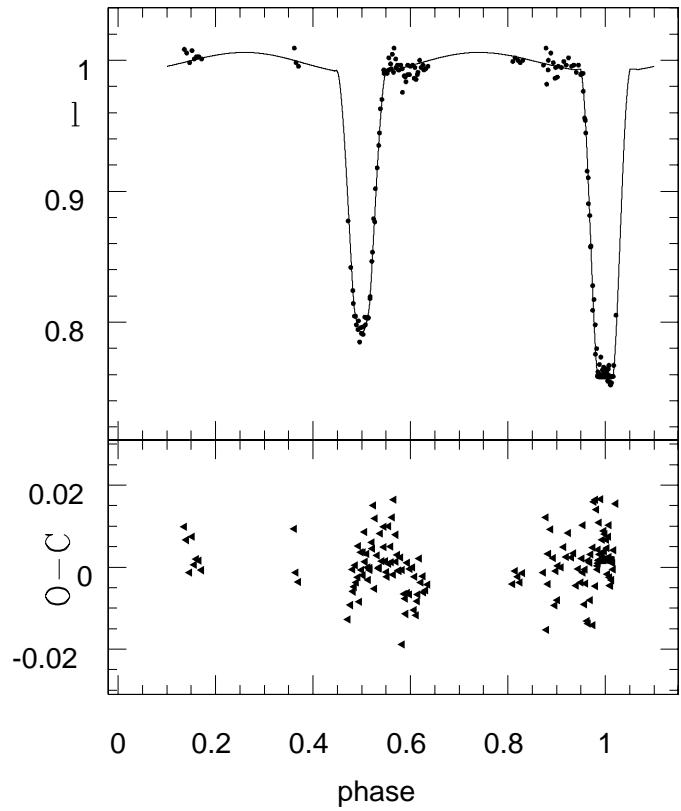}
\caption{The observed $V$-band light curve (in intensity units)
overplotted with the photometric solution. Residuals for the solution
in Mode 2 are shown at the bottom.  \label{fig5}}
\end{figure}
\clearpage
\begin{figure}
\figurenum{6}
\plotone{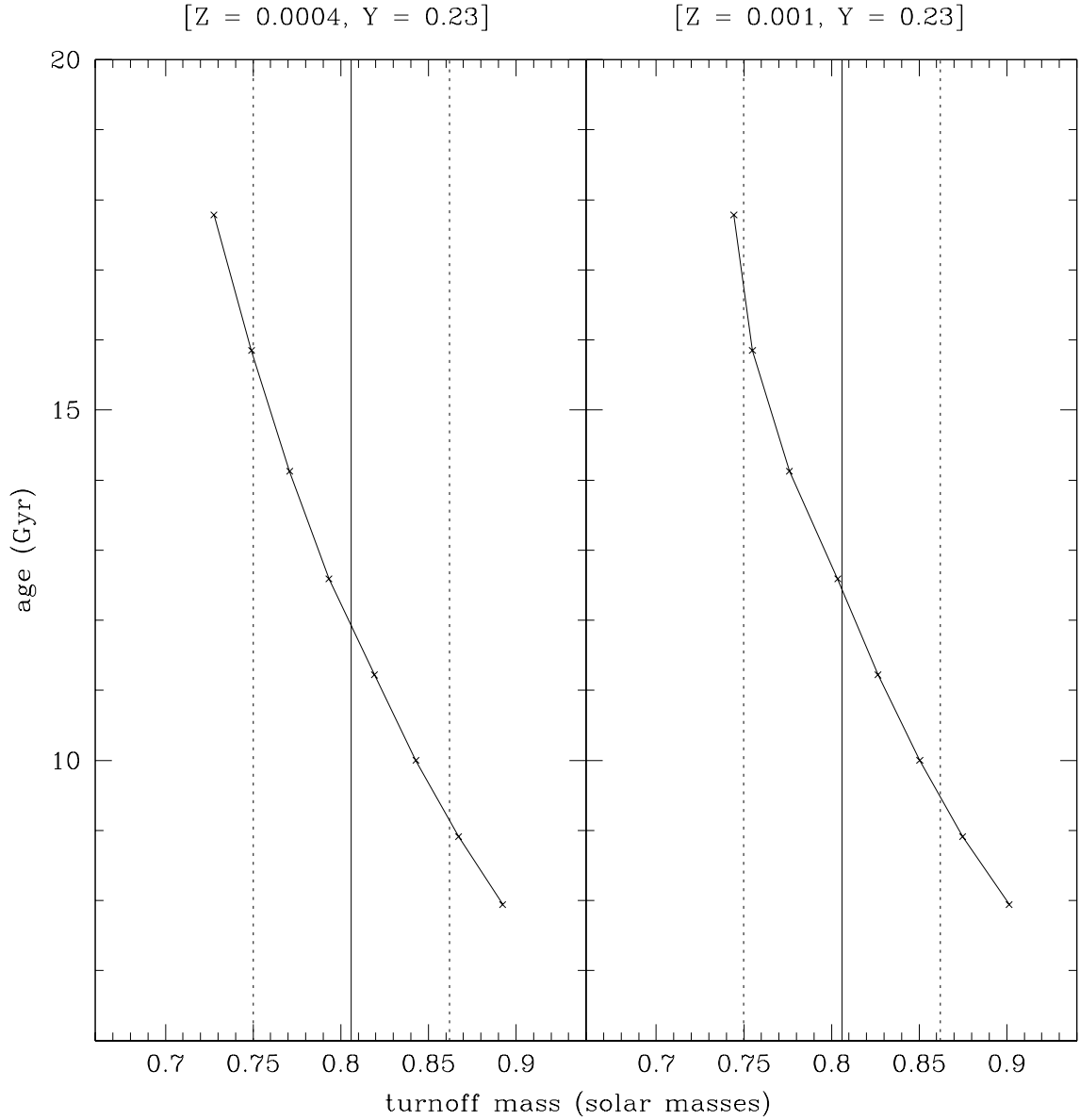}
\caption{Age - turnoff mass relations from \citet{gira00}. The left
panel is for Z = 0.0004, Y = 0.23 and the right panel is for Z = 0.001,
Y = 0.23.  The vertical solid line represents the mass of the primary
of OGLEGC17 and the vertical dotted lines represent one sigma errors in
the mass. \label{fig6}}
\end{figure}
\clearpage
\begin{figure}
\figurenum{7}
\plotone{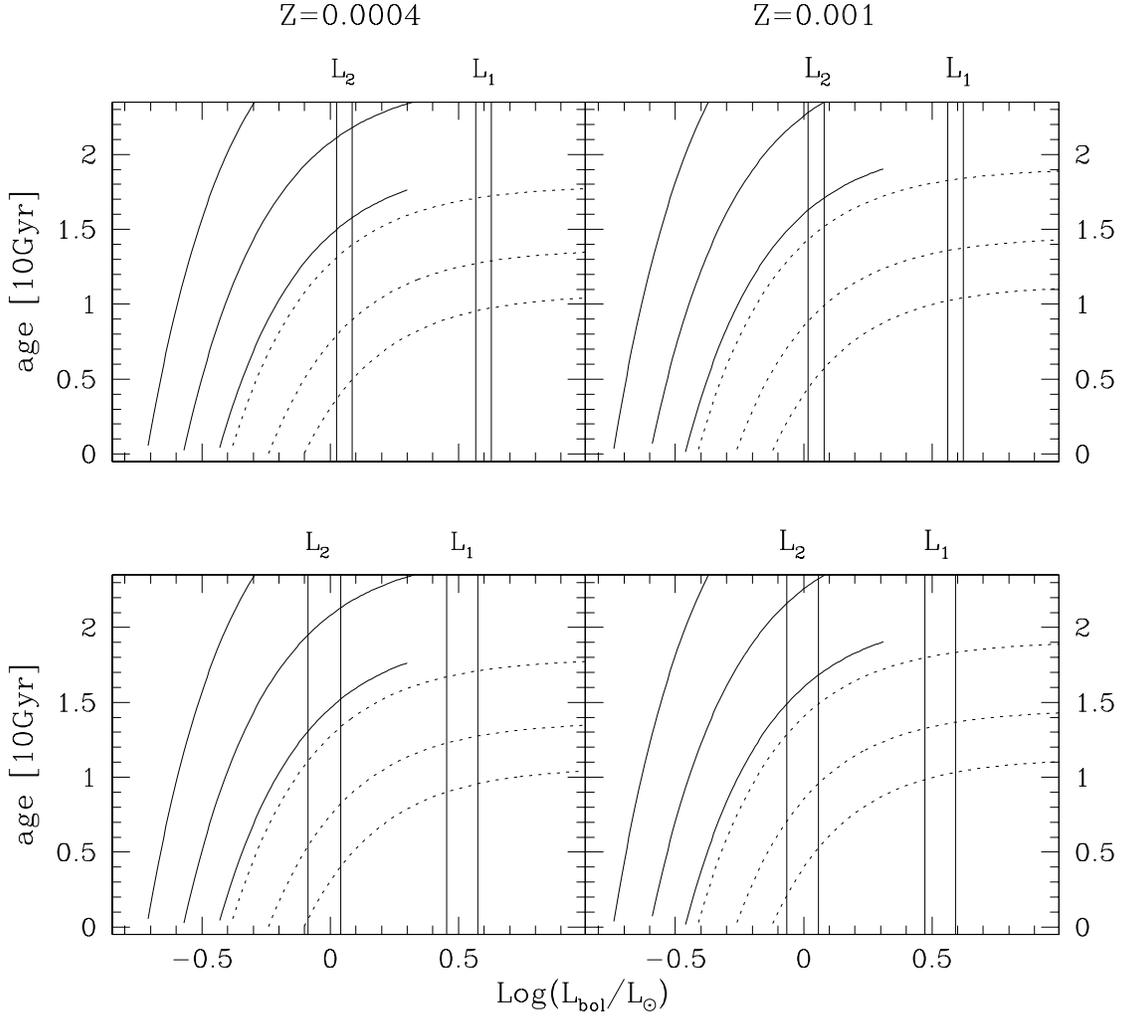}
\caption{Theoretical age-luminosity relations for masses
$m_{2}=0.686\pm 0.046M_{\odot}$ (continuous curves) and $m_1=0.806\pm
0.056M_{\odot}$ (dotted curves) from \citet{gira00}. Vertical lines
mark $\pm$one sigma ranges for the  observed luminosities of the
components of OGLEGC17. The upper panel corresponds to bolometric
luminosities derived with a $V-K$ calibration of effective
temperatures.  The lower panel corresponds to bolmtric luminosities
derived with a $B-V$ calibration. See text for details.  \label{fig7}}
\end{figure}
\clearpage
\begin{figure}
\figurenum{8}
\plotone{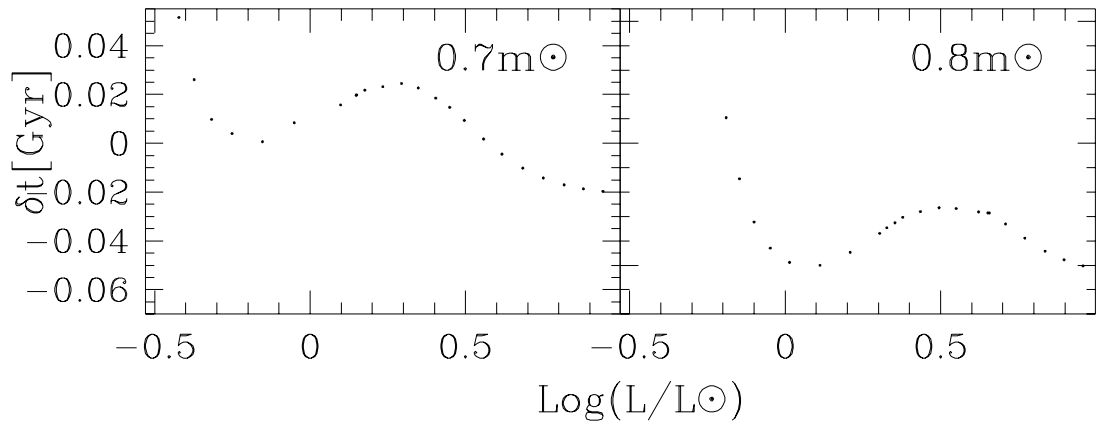}
\caption{Differences between age-luminosity relations from Girardi et
al.  (2000) and Weiis \& Schlattl (2000). Difference in age predicted
for a given luminosity are shown for masses $0.7m_{\odot}$  and
$0.8m_{\odot}$ and for models with Z=0.001 and Y=0.23. \label{fig8}}
\end{figure}
\clearpage
\begin{figure}
\figurenum{9}
\plotone{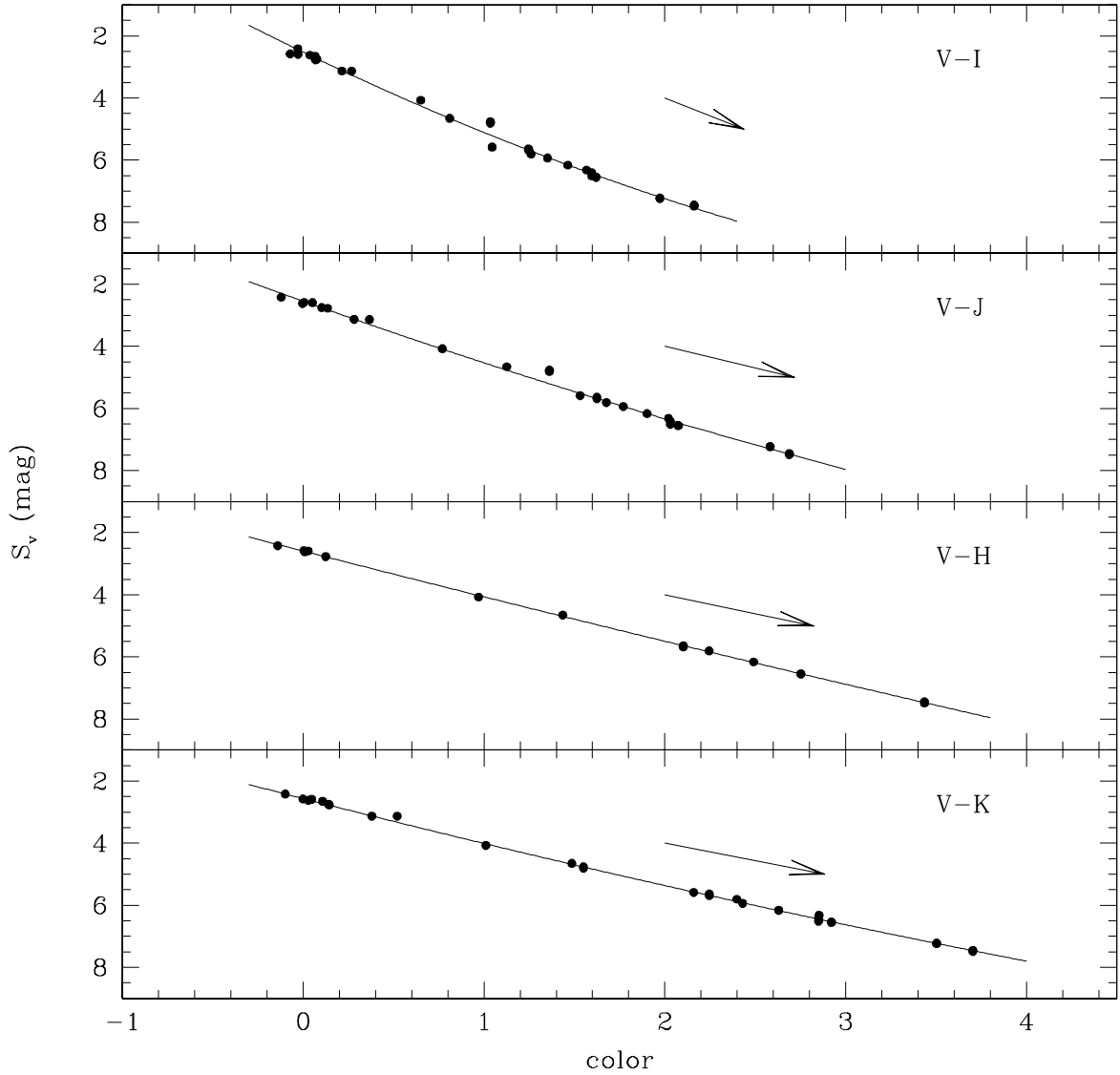}
\caption{The surface brightnesses of nearby stars in the $V$-band is
plotted against $V-I, V-J, V-H$, and $V-K$ colors. The solid lines are
quadratic least-squares fits to the data, the coefficients of the fits
are listed in Table 9.  The  arrows  correspond to a reddening $A_{\rm
V}$ of one magnitude.  \label{fig9}}
\end{figure}
\clearpage
\begin{figure}
\figurenum{10}
\plotone{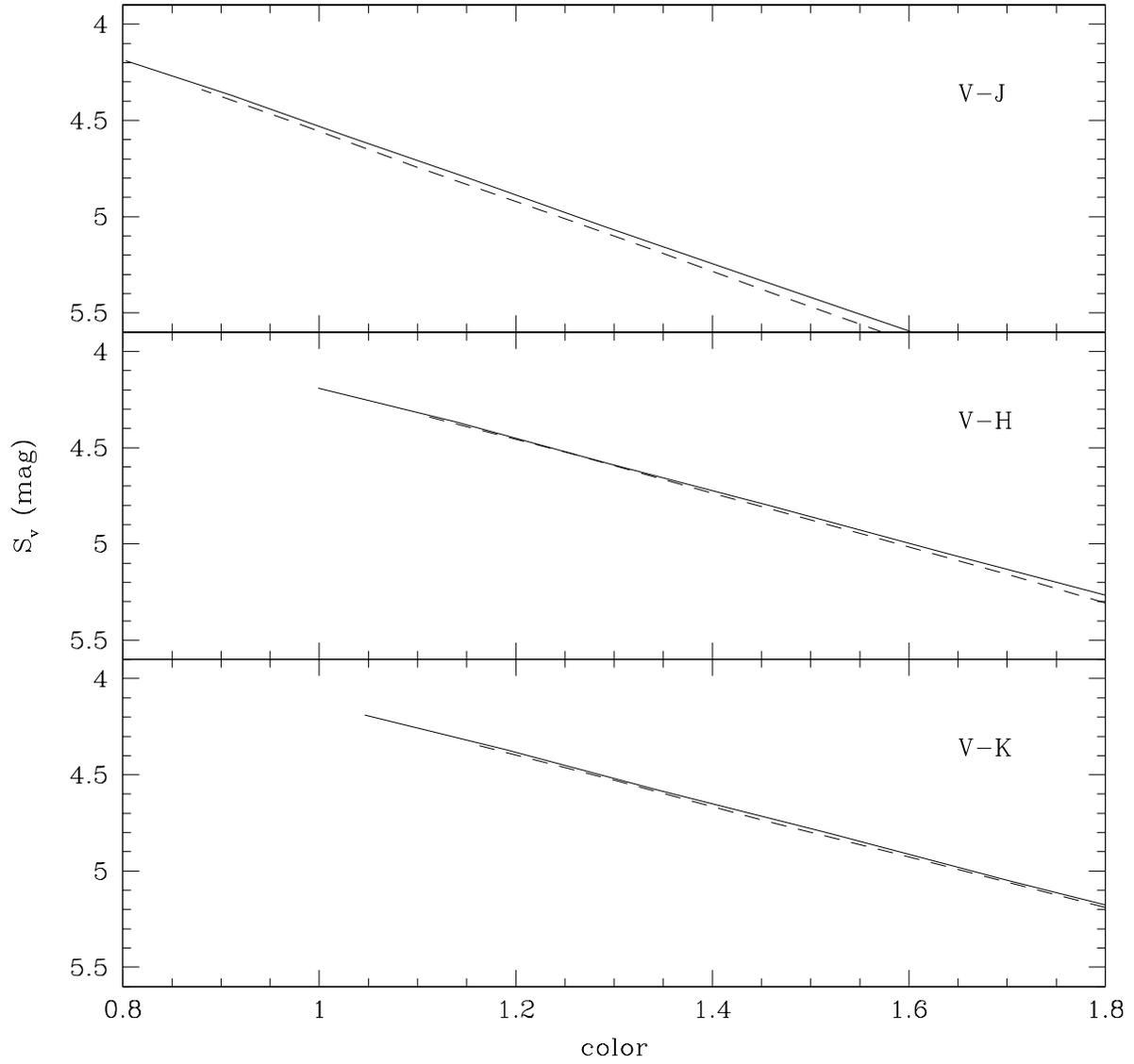}
\caption{Synthetic relations $S_{\rm V}$-$(V-K)$ for ${\rm [Fe/H]}=0$
(solid lines) and ${\rm [Fe/H]}=-2.0$ (dotted lines).  \label{fig10}}
\end{figure}
\clearpage
\begin{deluxetable}{llllll}
\small
\tablecaption{BVI photometry of OGLEGC17 at minima and at 
quadrature\label{tbl-1}}
\tablewidth{0pt}
\tablehead{
\colhead{phase} & \colhead{$V$} & \colhead{$B$} & \colhead{$I$} & \colhead{$B-V$}   & \colhead{$V-I$} \\
}
\startdata
Max & 17.155 & 17.790 & 16.340 & 0.635 & 0.815 \\
Min I & 17.450 & 18.110 & 16.63 & 0.66 & 0.82 \\
Min II & 17.41 & 18.050 & 16.58 & 0.640 & 0.83 \\
\enddata
\end{deluxetable}

\clearpage

\begin{deluxetable}{ccccccc}
\small
\tablecaption{Infrared photometry of OGLEGC17 \label{tbl-2}}
\tablewidth{0pt}
\tablehead{
\colhead{HJD\tablenotemark{a}} & \colhead{phase} & \colhead{$J$} & \colhead{phase} & \colhead{$H$} & \colhead{phase}   & \colhead{$K_s$}}

\startdata
09.849 & 0.403 & 15.82$\pm$0.02 & 0.396 & 15.47$\pm$0.02 & 0.392 & 15.42$\pm$0.02 \\
11.882 & 0.228 & 15.85$\pm$0.02 & 0.225 & 15.49$\pm$0.01 & 0.217 & 15.41$\pm$0.03 \\
12.877 & 0.631 & 15.90$\pm$0.01 & 0.627 & 15.50$\pm$0.01 & 0.617 & 15.48$\pm$0.02 \\
14.792 & 0.407 & 15.89$\pm$0.02 & 0.410 & 15.49$\pm$0.02 & 0.419 & 15.44$\pm$0.03 \\
16.872 & 0.250 & 15.86$\pm$0.02 & 0.246 & 15.48$\pm$0.03 & 0.233 & 15.46$\pm$0.02 \\ 
17.780 & 0.618 & 15.88$\pm$0.02 & 0.621 & 15.50$\pm$0.02 & 0.629 & 15.48$\pm$0.03 \\
\hline
average      &       & 15.867$\pm$0.029 &     & 15.488$\pm$0.012 &     & 15.448$\pm$0.030 \\
\enddata
\tablenotetext{a}{HJD for $J$-band observations - 2451200.0}
\end{deluxetable}

\clearpage

\begin{deluxetable}{cccccccc}
\small
\tablecaption{Velocity measurements for OGLEGC17 \label{tbl-3}}
\tablewidth{0pt}
\tablehead{
\colhead{Number} & \colhead{\# of spectra\tablenotemark{a}} & 
\colhead{$\phi$} & \colhead{$\Delta\phi$}  &
\colhead{$V_1$} & \colhead{$\sigma(V_1)$} &
\colhead{$V_2$} & \colhead{$\sigma(V_1)$}
}
\startdata
 1 & 3 & 0.001 & 0.078 & 232.9 &  9.7 & 232.9 &  9.7 \\ 
 2 & 5 & 0.142 & 0.030 & 305.8 & 15.4 & 167.7 & 20.7 \\  
 3 & 5 & 0.176 & 0.035 & 308.5 & 10.6 & 141.0 & 13.2 \\
 4 & 4 & 0.206 & 0.022 & 317.5 & 16.2 & 140.5 &  9.8 \\
 5 & 4 & 0.233 & 0.021 & 314.0 & 10.2 & 142.2 & 19.4 \\
 6 & 6 & 0.346 & 0.067 & 309.6 &  8.3 & 169.2 & 13.2 \\
 7 & 4 & 0.562 & 0.031 & 213.8 &  6.1 & 260.7 & 11.3 \\
 8 & 2 & 0.768 & 0.014 & 145.3 &  9.3 & 323.6 & 14.2 \\
 9 & 5 & 0.795 & 0.073 & 170.8 & 11.6 & 332.7 & 10.7 \\
10 & 5 & 0.039 & 0.042 & 283.5 & 13.9 & 233.6 & 15.2 \\
11 & 4 & 0.085 & 0.031 & 276.9 & 14.2 & 216.4 & 15.5 \\
12 & 6 & 0.247 & 0.056 & 314.0 & 14.6 & 134.3 & 10.9 \\
13 & 7 & 0.826 & 0.066 & 160.4 &  8.3 & 332.3 & 12.3 \\
\enddata

\tablenotetext{a}{Number of  spectra averaged}
\end{deluxetable}

\clearpage

\begin{deluxetable}{ll}
\small
\tablecaption{Elements for OGLEGC17 \label{tbl-4}}
\tablewidth{0pt}
\tablehead{
\colhead{Element} & \colhead{Value} 
}
\startdata

$P$~(days) & 2.4669384~$\pm$~0.0000023\tablenotemark{a} \\
$T_0$~(HJD) & 2449082.3530~$\pm$~0.0008\tablenotemark{a} \\
$K_1$~(km s$^{-1}$) & 82.65~$\pm$~3.25 \\
$K_2$~(km s$^{-1}$) & 97.04~$\pm$~3.89 \\
$\gamma$~(km s$^{-1}$) & 237.97~$\pm$~1.93 \\
$e$ & 0.0\tablenotemark{b} \\
$m_{2}/m_{1}$ & 0.852~$\pm$~0.073 \\
\enddata

\tablenotetext{a}{Photometric ephemeris}
\tablenotetext{b}{Assumed value}

\end{deluxetable}

\clearpage
\begin{deluxetable}{lll}
\small
\tablecaption{Photometric solutions of $BVI$ light curves 
of OGLEGC17 \label{tbl-5}}
\tablewidth{0pt}
\tablehead{
\colhead{parameter} & \colhead{Mode 0} & \colhead{Mode 2}  \\
}
\startdata
$i$ & 87.162 $\pm$1.98        & 86.31 $\pm$1.04 \\
$\Omega_1$ & 5.645 $\pm$ 0.171 & 5.562 $\pm$0.063\\
$\Omega_2$ & 9.745 $\pm$ 0.337 & 9.523 $\pm$0.133\\
$T_1$     & 5676\tablenotemark{*} & 5676\tablenotemark{*} \\
$T_2$     & 6085\tablenotemark{*} & 6077$\pm$ 40 \\ 
$(l1/(l1+l2)_{B}$ & 0.7597 $\pm$ 0.0029& 0.7608 $\pm$ 0.0081\\
$(l1/(l1+l2)_{V}$ & 0.7725 $\pm$ 0.0028& 0.7694 $\pm$ 0.0078\\
$(l1/(l1+l2)_{I}$ & 0.7811 $\pm$ 0.0039& 0.7855 $\pm$ 0.0111\\
$(l2/(l1+l2)_{B}$ & 0.2403 $\pm$ 0.0034& - \\
$(l2/(l1+l2)_{V}$ & 0.2275 $\pm$ 0.0033& - \\
$(l2/(l1+l2)_{I}$ & 0.2189 $\pm$ 0.0053& - \\
$r_{1}$(pole) & 0.2079$\pm$ 0.0073& 0.2114$\pm$ 0.0028\\
$r_{1}$(point)& 0.2125$\pm$ 0.0080& 0.2166$\pm$ 0.0031\\
$r_{1}$(side) & 0.2096$\pm$ 0.0076& 0.2133$\pm$ 0.0029\\
$r_{1}$(back) & 0.2119$\pm$ 0.0079& 0.2157$\pm$ 0.0031\\
$r_{2}$(pole) & 0.0982$\pm$ 0.0038& 0.1007$\pm$ 0.0016\\
$r_{2}$(point)& 0.0984$\pm$ 0.0038& 0.1010$\pm$ 0.0016\\
$r_{2}$(side) & 0.0983$\pm$ 0.0038& 0.1009$\pm$ 0.0016\\
$r_{2}$(back) & 0.0984$\pm$ 0.0038& 0.1010$\pm$ 0.0016\\
$<r_1>$       & 0.2105$\pm$ 0.0077& 0.2142$\pm$ 0.0030 \\
$<r_2>$       & 0.0983$\pm$ 0.0038& 0.1009$\pm$ 0.0016 \\

\enddata

\tablenotetext{*}{not adjusted}

\end{deluxetable}
\clearpage

\begin{deluxetable}{ll}
\small
\tablecaption{Absolute parameters for OGLEGC17 \label{tbl-6}}
\tablewidth{0pt}
\tablehead{
\colhead{Element} & \colhead{Value} 
}
\startdata
$A$~($R_{\odot}$) & 8.782~$\pm$~0.248 \\
$R_1$~($R_{\odot}$) & 1.882~$\pm$~0.059 \\
$R_2$~($R_{\odot}$) & 0.886~$\pm$~0.029 \\
$M_1$~($M_{\odot}$) & 0.806~$\pm$~0.056 \\
$M_2$~($M_{\odot}$) & 0.686~$\pm$~0.047 \\
\enddata

\end{deluxetable}

\clearpage

\begin{deluxetable}{llllll}
\small
\tablecaption{${Teff}$ and $L_{bol}$ for components of OGLEGC17 \label{tbl-7}}
\tablewidth{0pt}
\tablehead{
\colhead{ ${\rm [Fe/H]}$} & \colhead{color} & \colhead{$Teff_{1}$} & \colhead{$Teff_{2}$}& \colhead{ $L_{1}/L_{\odot}$}& \colhead{ $L_{2}/L_{\odot}$} 
}
\startdata
$-1.74$ & $B-V$ & $5676\pm 176$  & $6074\pm 192$ &$3.30\pm 0.46$ & $0.96\pm 0.14$ \\
$-1.33$ & $B-V$ & $5739\pm 176$  & $6141\pm 192$ &$3.44\pm 0.47$ & $1.00\pm 0.14$ \\
$-1.74$ & $V-K$ & $5948\pm 44$  & $6341\pm 47$ &$3.97\pm 0.28$ & $1.14\pm 0.08$ \\
$-1.33$ & $V-K$ & $5926\pm 44$  & $6322\pm 47$ &$3.92\pm 0.27$ & $1.12\pm 0.08$ \\
\enddata

\end{deluxetable}

\clearpage

\begin{deluxetable}{llllll}
\small
\tablecaption{Absolute visual magnitudes and distance moduli 
for components of OGLEGC17 
quadrature\label{tbl-8}}
\tablewidth{0pt}
\tablehead{
\colhead{$[Fe/H]$}&\colhead{color} & \colhead{$M_{\rm V1}$}&\colhead{$M_{\rm V2}$}&
\colhead{$(m-M)_{\rm V1}$}&\colhead{$(m-M)_{\rm V2}$} \\
}
\startdata
-1.74 & $B-V$ & $3.65\pm0.16$ &$4.96\pm0.17$ &$13.81\pm0.16$ &$13.76\pm0.17$ \\
-1.33 & $B-V$ & $3.56\pm0.16$ &$4.89\pm0.17$ &$13.90\pm0.16$ &$13.83\pm0.17$ \\
-1.74 & $V-K$ & $3.42\pm0.09$ &$4.77\pm0.09$ &$14.04\pm0.09$ &$13.95\pm0.09$ \\
-1.33 & $V-K$ & $3.42\pm0.09$ &$4.76\pm0.09$ &$14.04\pm0.09$ &$13.96\pm0.09$ \\
\enddata
\end{deluxetable}

\clearpage

\begin{deluxetable}{ccccc}
\small
\tablecaption{Coefficients for $S_V$ - Color Relations \label{tbl-9}}
\tablewidth{0pt}
\tablehead{
\colhead{color} & \colhead{constant} & 
\colhead{linear} & \colhead{quadratic} & \colhead{$rms$} 

}
\startdata
$V-I$ & 2.522 & 2.816 & -0.228 & 0.072 \\
$V-J$ & 2.545 & 2.073 & -0.089 & 0.058 \\
$V-H$ & 2.592 & 1.496 & -0.022 & 0.028 \\
$V-K$ & 2.563 & 1.493 & -0.046 & 0.023 \\
\enddata

\end{deluxetable}

\clearpage

\begin{deluxetable}{cccc}
\small
\tablecaption{Colors for OGLEGC17 \label{tbl-10}}
\tablewidth{0pt}
\tablehead{
\colhead{color} & \colhead{Max} & 
\colhead{Primary} & \colhead{Secondary}  
}
\startdata
$(V-J)_0$ & 0.999 & 1.017 & 0.844  \\
$(V-H)_0$ & 1.336 & 1.369 & 1.163  \\
$(V-K)_0$ & 1.350 & 1.400 & 1.167  \\
\enddata

\end{deluxetable}

\clearpage

\begin{deluxetable}{ccccccc}
\small
\tablecaption{Surface Brightnesses and Distances for OGLEGC17 \label{tbl-11}}
\tablewidth{0pt}
\tablehead{
\colhead{color} &  
\colhead{$S_{V,pri}$} & \colhead{$S_{V,sec}(pc)$} & 
\colhead{$d_{pri}(pc)$} & \colhead{$d_{sec}(pc)$} & 
\colhead{$(m-M)_{V,pri}$} & \colhead{$(m-M)_{V,sec}$} 
}
\startdata
$(V-J)$ & 4.617 & 4.271 & 5352 & 5385 & 14.04 & 14.06 \\
$(V-H)$ & 4.624 & 4.322 & 5333 & 5259 & 14.04 & 14.01 \\
$(V-K)$ & 4.583 & 4.262 & 5435 & 5406 & 14.08 & 14.07 \\
\enddata

\end{deluxetable}


\begin{thebibliography}{}

\bibitem[Alonso et al. (1996)]{alonso96} Alonso, A., Arribas, S., \&
Martinez-Roger 1996, \aap, 313, 873
\bibitem[Alonso et al. (1999)]{alonso99} Alonso, A., Arribas, S., \&
Martinez-Roger 1999, \aaps, 140, 261
\bibitem[Anderson (1991)]{and91} Anderson, J. 1991, \aapr, 3, 91
\bibitem[Barnes and Evans (1976)]{barnes76} Barnes, T. G., and Evans,
D. S. 1976, \mnras, 174, 489
\bibitem[Bell et al. (1993)]{bell93} Bell, S. A., Hill, G., Hilditch,
R. W., Clausen, J. V., and Reynolds, A. P. 1993, \mnras, 265, 1047
\bibitem[Bertelli et al. (1994)]{bert94} Bertelli, G., Bressan, A.,
Chiosi, C., Fagotto, F., \& Nasi, E. 1994, \aaps, 106, 275 
\bibitem[Bessel and Brett (1988)]{bessel88} Bessel, M. S. \& Brett, J. M.
1988, \pasp, 100, 1134
\bibitem[Claret (1995)]{clar95} Claret, A., Diaz-Cordoves, J., \&
Gimenez, A. 1995, \aaps, 114, 247 
\bibitem[Diaz-Cordovez et al. (1995)]{diaz95} Diaz-Cordoves, 
J., Claret, A., \& Gimenez, A. 1995, \aap, 110, 329
\bibitem[Di Benedetto (1998)]{dib98} Di Benedetto, G. P. 1998, \aap,
339, 858
\bibitem[Fitzpatrick et al. (2000)]{fitz01} Fitzpatrick, E. L., Ribas, I.,
Guinan, E. F., DeWarf, L. E., Maloney, F. P., \& Massa, D. 2001, \apj,
in press
\bibitem[Girardi et al. (2000)]{gira00} Girardi, L., Bressan, A., Bertelli,
G., \& Chiosi, C. 2000, \aaps, 141, 371
\bibitem[Guinan et al. (1998)]{guinan98} Guinan, E. F. et al. 1998,
\apjl, 509, L21
\bibitem[Hajian et al. (1998)]{haj98} Hajian, A. R. et al. 1998,
\apj, 496, 484
\bibitem[Houdashelt et al. (2000)]{houd00}Houdashelt, M.J., Bell, R.A.,
\& Sweigert, A. V. 2000, \aj, 119, 1448
\bibitem[Harris (1996)]{harris96} Harris, W. E. 1996, \aj, 112, 1487
\bibitem[Hughes and Wallerstein (2000)]{hughes00} Hughes, J., \& 
Wallerstein, G. 2000, \aj, 119, 1225
\bibitem[Kaluzny et al. (1996)]{jka96} Kaluzny, J. K., Kubiak, M.,
Szyma\'nski, M., Udalski, A., Krzemi\'nski, and Mateo, M. 1996, \aaps, 
120, 139
\bibitem[Kaluzny et al. (1998)]{jka98} Kaluzny, J. K., Wysocka, A., 
Stanek, K., and Krzemi\'nski, W. 1998, Acta Astronomica, 48, 439
\bibitem[Kruszewski and Semeniuk (1999)]{krus99} Kruszewski, A. \&
Semeniuk, I. 1999, Acta Astronomica, 49, 561
\bibitem[Lacy (1977)]{lacy77} Lacy, C. H. 1977, \apj, 213, 458
\bibitem[Lacy (1979)]{lacy79} Lacy, C. H. 1979, \apj, 228, 817
\bibitem[Landolt (1992)]{land92} Landolt, A. 1992, \aj, 104, 340
\bibitem[Latham et al. (1988)]{lath88} Latham, D. W., Mazah, T.,
Carney, B. W., McCroskey, R. E., Stefanik, R. P., and Davis, R. J.
1988, \aj, 96, 567
\bibitem[Lee (1999)]{lee00} Lee, Y.-W., Joo, J.-M, Sohn, Y-J., Rey,
S.-C., Lee, H.-C., \& Walker, A.R. 1999, Nature, 402, 55L
\bibitem[Leung and Wilson (1977)]{leun77} Leung, K.-C., \& Wilson, R.E.
1977, \apj, 211, 853
\bibitem[Lucy and Sweeny (1971)]{lucy71} Lucy, L. B. and Sweeny, M. A. 1971,
\aj, 76, 544
\bibitem[Majewski (2000)]{maj00} Majewski, S.R., Petterson, R.J.,
Dinescu, D.I., Johnson, W.Y., Ostheimer, J.C., Kunkel, W.E., \& Palma,
C. 2000, in The Galactic Halo: From Globular Clusters to Field Stars,
ed. A.Noels et al. (Liege: Univ. Liege, Inst. d'Astrophys.Geophys.), in press
; /astro-ph/9910278
\bibitem[Metcalfe (1996)]{metcalfe96} Metcalfe, T.S., Mathieu, R.D.,
Latham, D.W., \& Torres, G. 1996, \apj, 456, 356
\bibitem[Norris (1975)]{norr75} Norris, J., \& Bessell, M.S. 1975, \apj,
211, L91
\bibitem[Paczy\'nski (1997)]{pac97} Paczy\'nski, B.  1997, The Extragalactic Distance Scale STScI Symposium,
(Cambridge: Cambridge University Press)
\bibitem[Perryman et al. (1997)]{perry97} Perryman, M. A. C. et al. 1997, \aap,
323, L49
\bibitem[Persson et al. (1992)]{sep92} Persson, S. E., West, S. C.,
Carr, D. M., Sivaramakrishnan, A., and Murphy, D. C. 1992, \pasp,
104, 204
\bibitem[Persson et al. (1998)]{sep98} Persson, S. E., Murphy, D. C.,
Krzeminski, W., Roth, M., and Rieke, M. J. 1998, \aj, 116, 2475
\bibitem[Plewa (1988)]{plewa98} Plewa, T. 1988, Acta Astron., 38, 415
\bibitem[Reid (1998)]{reid98} Reid, I. N. 1998, \aj, 115, 204
\bibitem[Renzini (1991)]{renzini91} Renzini, A. 1991 in Observational
Tests of Cosmological Inflation, edited by T. Shanks et al. (Dordrecht: 
Kluwer), 131
\bibitem[Rieke and Lebofsky (1985)]{rieke85} Rieke, G., \& Lebofsky, M.
1985, \apj, 288, 618
\bibitem[Salaris (1998)]{salaris00} Salaris, M., \& Weiss, A. 1998,
\aap, 335, 943
\bibitem[Schechter, Mateo, and Saha (1993)]{sch93} Schechter, P. S.,
Mateo, M., and Saha, A. 1993, \pasp, 105,1342
\bibitem[Shectman (1984)]{shec84} Shectman, S. A. 1984, Instrumentation
in Astronomy V, Proc SPIE, edited by A. Boksenberg and D. L. Crawford
(SPIE, Bellingham), 445, 128
\bibitem[Schlegel (1998)]{schl98} Schlegel, D.J., Finkbeiner, D.P., 
Davis, M. 1998, \apj, 500, 525
\bibitem[Semeniuk (2000)]{sem00} Semeniuk, I. 2000, Acta Astronomica, 50, 381
\bibitem[Sekiguchi \& Fukugita (2000)]{seki00} Sekiguchi, M., \& Fukugita, M. 2000, \aj, 120, 1072
\bibitem[Stebbins (1910)]{stebbins10} Stebbins, J. 1910, \apj, 32, 185
\bibitem[Stetson (1987)]{stet87} Stetson, P.B. 1987, \pasp, 99, 191
\bibitem[Stetson (1990)]{stet90} Stetson, P.B. 1990, \pasp, 102, 932
\bibitem[Suntzeff (1996)]{suntz96} Suntzeff, N.B., \& Kraft, R.P.  1996, \aj, 111, 1913
\bibitem[Tokunaga (2000)]{tokunaga00} Tokunaga, A.T. 2000, in Allens's
Astrophysical Quantities, ed. A.N. Cox, p. 143 (Springer-Verlag)
\bibitem[Tomkin et al. (1992)]{tomkin} Tomkin, J., Lemke, M., Lambert,
D. L., and Sneden, C. 1992, \aj, 104, 1568
\bibitem[Torres et al. (1997)]{torres97} Torres, G., Stefanik, R.P.,
Andersen, J., Nordstrom, B., Latham, D., \& Clausen, J.V. 1997,
\aj, 114, 2764
\bibitem[VandenBerg, Bolte, and Stetson (1996)]{vdb96} VandenBerg, D. A.,
Bolte, M., and Stetson, P. B. 1996, \araa, 34, 461
\bibitem[Weiss (2000)]{weiss00} Weiss, A., \& Schlattl, H. 2000,
\aaps, 144, 487
\bibitem[Wilson (1979)]{wil79} Wilson, R.E. 1979, \apj, 234, 1054
\bibitem[Wilson (1971)]{wil72} Wilson, R.E., \& Devinney, E.J. 1971, 
\apj, 166, 605

\end{thebibliography}
\end{document}